\documentclass[12pt]{article}

\usepackage{times,amsmath,epsfig}
\usepackage{graphicx}
\usepackage[ruled,oldcommands]{algorithm2e}

\newcommand{\vs}{\vspace{1ex}}		
\newcommand{\vsp}{\vspace{2ex}}		

\newcommand{\negvs}{\vspace*{-1ex}}	

\newcommand{\hs}{\hspace*{1em}}		
\newcommand{\hsp}{\hspace*{2em}}	

\newcommand{\ie}{\mbox {\em i.e.}}

\newcommand{\nop}[1]{}		

\newtheorem{theorem}{Theorem}
\newtheorem{lemma}{Lemma}
\newtheorem{definition}{Definition}
\newtheorem{example}{Example}

\def\blackbox{{\rule{2mm}{2mm}}}
\def\qed{\hspace*{\fill}\blackbox}

\newcommand{\examplebox}{\qed}

\title{Integration of Probabilistic Uncertain Information}

\author{Fereidoon Sadri and Gayatri Tallur\\
\\
Department of Computer Science\\
University of North Carolina at Greensboro\\
{\small \tt f\_sadri@uncg.edu, gtallur@gmail.com}
}

\date{}

\begin{document}
\maketitle

\begin{abstract}

We study the problem of data integration from sources that contain probabilistic uncertain information. Data is modeled by possible-worlds with probability distribution, compactly represented in the probabilistic relation model. 
Integration is achieved efficiently using the extended 
probabilistic relation model. 
We study the problem of determining the probability distribution of the integration result. It has been shown that, in general, only probability ranges can be determined for the result of integration. In this paper we concentrate on a subclass of extended probabilistic relations, those that are obtainable through integration. We show that under intuitive and reasonable assumptions we can determine the exact probability distribution of the result of integration.  
\end{abstract}

\section{Introduction}
\label{sec:intro}

Information integration and modeling and management of uncertain information
have been active research areas for decades, with both areas receiving
significant renewed interest in recent years
\cite{Antova-Jansen-Koch-Olteanu:icde08,
Antova-Koch-Olteanu:icde07,
Chen-Chirkova-Sadri-Salo:actainformatica13,
Dalvi-Re-Suciu:cacm09,
Haas:icdt07,
Halevy-Rajaraman-Ordille:vldb06}.
The importance of information integration {\em with uncertainty},
on the other hand, has been realized more recently
\cite{Dong-Halevy-Yu:vldb07,
Dong-Halevy-Yu:vldbj09,
Eshmawi-Sadri:ideas09,
Haas:icdt07,
Halevy-etal:sigmod05,
Halevy-Rajaraman-Ordille:vldb06,
Magnani-Montesi:mud07,
Magnani-Montesi:jdiq10,
Olteanu-Huang-Koch:icde09,
Re-Dalvi-Suciu:icde07,
Sarma-et-al:vldbj09,
Sarma-Benjelloun-Halevy-Widom:icde06,
Sen-Deshpande:icde07}. 
It has been observed that
\cite{Halevy-Rajaraman-Ordille:vldb06}
``While in traditional database management managing uncertainty and lineage seems like a nice
feature, in data integration it becomes a necessity.''

The widely accepted conceptual model for uncertain data is 
the possible-worlds model
\cite{Abiteboul-Kanellakis-Grahne:sigmod87}.
For practical applications, a representation of choice 
is the probabilistic relation model 
\cite{Dalvi-Suciu:vldb04,Dalvi-Suciu:vldbj07}, which provides a compact and efficient representation for uncertain data. 
We have shown that integration of uncertain data
represented in the probabilistic relation model can be
achieved efficiently using the extended probabilistic relation
model \cite{Borhanian-Sadri:ideas13}.
 
In this paper we concentrate on the integration of
{\em probabilistic} uncertain data. 
We study the problem of determining the 
probability distribution of the integration result.
A recent work has shown how to obtain 
probability {\em ranges} for the result of integration 
\cite{Sadri:cikm12}.
We study this problem in two frameworks: The probabilistic 
possible-worlds model, and the probabilistic relation model. 
We show that, under intuitive and reasonable assumptions, 
we can determine the exact probability distribution 
of integration in either of the frameworks. 
Further, we show that the two approaches are equivalent 
while the probabilistic relation approach provides a significantly more efficient method in practice.

\vs
We make the following contributions

\begin{itemize}

\item
We review the integration problem in the probabilistic 
possible-worlds model, and show why, in the general case, 
it is only possible to determine probability ranges for
the integration result.

\item
We add an intuitive an realistic assumption regarding
the probabilistic correlation of the inputs, and
show that under this assumption exact probability 
distribution can be obtained for the integration result.

\item
We concentrate on the integration problem in the
probabilistic relation framework. We show that adding
an intuitive and realistic assumption in this framework
makes it possible to determine exact probability distribution 
for the integration.

\item
We show that the two approaches are equivalent in the following sense. First, the assumptions in the two frameworks, 
although different in appearence, are indeed equivalent.
Second, given the same inputs, the probability 
distributions obtained in the two approaches are the same.
This equivalence is a strong justification of the
robustness of our approaches.

\end{itemize}

This paper is organized as follows: We summarize some of the important concepts and results from 
\cite{Borhanian-Sadri:ideas13,Sadri:cikm12} 
in Section~\ref{sec:prelim}, and discuss the problem of integrating probabilistic data in Section~\ref{sec:prob-integ}. 
{\em Integrated Extended Probabilistic Relations} are introduced in Section~\ref{sec:pr-framework}. 
We study their properties, and present algorithms for determining if an epr-relation is the result of data integration.
Section~\ref{sec:prob-computation} is devoted to the discussion of computing probability distribution for the result of an integration. We present two approaches, and show they are equivalent. This is a further justification of our probability computation solutions. Conclusions are presented in 
Section~\ref{sec:conclusion}.

\section{Preliminaries}
\label{sec:prelim}

Foundations of uncertain information integration were discussed in the seminal work of 
Agrawal {\em et al} \cite{Agrawal-DasSarma-Ullman-Widom:vldb10}.
They discuss the fundamental concept of {\em containment} for uncertain databases,
and introduce alternative formulations for {\em equality} and {\em superset containment}.
Equality containment integration is more restrictive and applies to
cases where each information source has access only to
a portion of an uncertain database that is existing but unknown.
Superset containment integration is applicable in settings where we have uncertain data about
the real world from multiple sources and wish to integrate the data to obtain
the real world. The goal of integration is to obtain the best possible uncertain database that
contains all the information implied by sources, and nothing more.
An alternative formulation to superset-containment-based integration was presented in \cite{Sadri:cikm12}.
These approaches are based on the well-known {\em possible-worlds} model
of uncertain information \cite{Abiteboul-Kanellakis-Grahne:sigmod87}.
The possible-worlds model is widely accepted as the conceptual model
for uncertain information, and is used as the theoretical basis for 
operations and algorithms on uncertain data. But it is not, in general,
a suitable representation for the {\em implementation} of uncertain
information systems due to lack of efficiency. Instead, compact representations,
such as the {\em probabilistic relation model} 
\cite{Dalvi-Suciu:vldb04,Dalvi-Suciu:vldbj07},
are more appropriate for the implementation.
The problem of integration of information represented by
probabilistic relations has been studied in 
\cite{Borhanian-Sadri:ideas13}, which presents
efficient algorithms for the integration.
In this section, we will review some of the observations and results from these works.

Let us begin with the following definition of {\em uncertain database}
from \cite{Agrawal-DasSarma-Ullman-Widom:vldb10}.

\begin{definition}
\label{def:uncertainDB}
An {\em uncertain database} $U$ consists of 
a finite set of tuples $T(U)$ and a nonempty set of 
possible worlds $PW(U) = \{D_1, \ldots, D_n\}$, 
where each $D_i \subseteq T(U)$
is a certain database.
\end{definition}

This definition adds tuple-set $T(U)$ to the 
possible-worlds model.
In fact, there may be tuples in the tuple set
that do not appear in any possible world
of the uncertain database $U$.
If $T(U)$ is not provided explicitly, then we use the set of all tuples in the possible worlds as the tuple set,
\ie, $T(U) = D_1 \cup \cdots \cup D_n$.
It is interesting to notice that this model exhibits both closed-world and open-world
properties: If a tuple $t \in T(U)$ does not appear in a possible world $D_i$, then
it is assumed to be {\em false} for $D_i$ (hence, closed-world assumption).
In other words, $D_i$ explicitly rules out $t$.
The justification is that the source providing the uncertain information represented by $U$
is aware of (the information represented by) all $t \in T(U)$.
If some $t \in T(U)$ is absent from $D_i$, then the source explicitly rules out $t$
from $D_i$.
On the other hand, a tuple $t \not \in T(U)$ is assumed possible for possible-worlds
$D_i$ (hence, open-world assumption). This distinction is important for integration:
Consider integrating $D_i$ from one source
with a possible-world $D'_j$ from another source.
Let a tuple $t \in D'j$ where $t \not \in D_i$. 
For the first case ($t \in T(U)$), $D_i$ and $D'_j$ are not compatible
and can not be integrated. This is because $D_i$ explicitly rules out $t$ while $D'_j$ explicitly includes it.
On the other hand, for the second case ($t \not \in T(U)$), $D_i$ and $D'_j$ can be integrated since
$D_i$ can accept $t$ as a valid tuple.
The following example from \cite{Sadri:cikm12} demonstrates the above observations.

\begin{figure}[h]
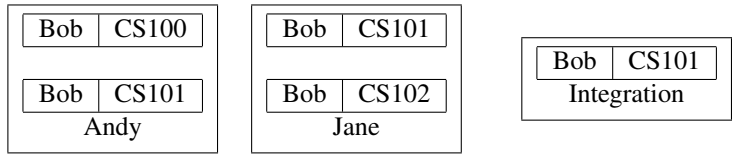

{\footnotesize
\begin{center}
\fbox{
\begin{tabular}{|c|c|} \hline
Bob     & CS100 \\ \hline
\multicolumn{2}{c}{} \\ \hline
Bob     & CS101 \\ \hline
\multicolumn{2}{c}{Andy}
\end{tabular}
}
\hs
\fbox{
\begin{tabular}{|c|c|} \hline
Bob     & CS101 \\ \hline
\multicolumn{2}{c}{} \\ \hline
Bob     & CS102 \\ \hline 
\multicolumn{2}{c}{Jane}
\end{tabular}
}
\hsp
\fbox{
\begin{tabular}{|c|c|} \hline
Bob     & CS101 \\ \hline
\multicolumn{2}{c}{Integration}
\end{tabular}
}
\end{center}
\caption{Possible-world relations of sources S1 (Andy), S2 (Jane), and integration result (Case 1)}
\label{fig:andy-jane-1}
} 
\end{figure}

\begin{example}
\label{ex:scenario2}
Andy and Jane are talking about fellow student Bob.
Andy says ``I am taking three courses, CS100, CS101, and CS102,
and Bob is in one of CS100 or CS101 (but not both).''
Jane says ``I am taking CS101 and CSC102 and Bob is
in one of them (but not both).''
These statements are represented by the possible-world relations
shown in Figure~\ref{fig:andy-jane-1}.
But Andy's tuple-set contains (Bob, CS102) hence, his statement 
also implies that Bob can not be in CS102.
So the result of integration is that Bob is taking CS101,
shown in Figure~\ref{fig:andy-jane-1}.
\end{example}

\begin{figure}[h]
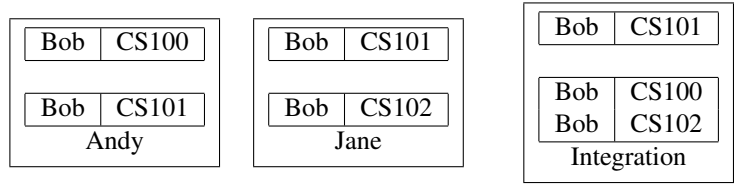

{\footnotesize
\begin{center}
\fbox{
\begin{tabular}{|c|c|} \hline
Bob     & CS100 \\ \hline
\multicolumn{2}{c}{} \\ \hline
Bob     & CS101 \\ \hline
\multicolumn{2}{c}{Andy}
\end{tabular}
}
\hs
\fbox{
\begin{tabular}{|c|c|} \hline
Bob     & CS101 \\ \hline
\multicolumn{2}{c}{} \\ \hline
Bob     & CS102 \\ \hline 
\multicolumn{2}{c}{Jane}
\end{tabular}
}
\hsp
\fbox{
\begin{tabular}{|c|c|} \hline
Bob     & CS101 \\ \hline
\multicolumn{2}{c}{} \\ \hline
Bob     & CS100 \\ 
Bob     & CS102 \\ \hline 
\multicolumn{2}{c}{Integration}
\end{tabular}
}
\end{center}
\caption{Possible-world relations of sources S1 (Andy), S2 (Jane), and integration result (Case 2)}
\label{fig:andy-jane-2}
} 
\end{figure}

Note that if Andy's tuple set did not contain (Bob, CS102),
i.e., if he was taking only CS100 and CS101 and had noticed Bob in one of them,
then his possible-world relations would still be the same.
But the result of integration in this case would contain a second possibility
that Bob is taking both CS100 and CS102.
This case is shown in Figure~\ref{fig:andy-jane-2}.

\subsection{Integration Algorithm for Uncertain Data Represented 
in the Possible-Worlds Model}
\label{sec:integ-pw}

Let $S$ and $S'$ be information sources
with possible worlds
$\{D_1,\ldots,D_n\}$ and $\{D'_1,\ldots,D'_{n'}\}$, respectively.
Let $T$ and $T'$ be the tuple-sets of $S$ and $S'$.
We need the following definition.

\begin{definition}
\label{def:compatible}
A pair of possible-world relations $D_i$ and $D'_j$ are
{\em compatible} if for each tuple $t \in T \cap T'$ either
both $D_i$ and $D_j$ contain $t$ (\ie, $t \in D_i$ and $t \in D'_j$), or
neither $D_i$ nor $D'_j$ contain $t$ (\ie, $t \not \in D_i$ and $t \not \in D'_j$).
Otherwise $D_i$ and $D'_j$ are not compatible.
\end{definition} 

Given information sources $S$ and $S'$, the integration algorithm (Algorithm~\ref{alg:integ-pw}) considers all 
possible-world pairs from the two sources.
If they are compatible, their union forms a possible-world of the integration.

\begin{algorithm}
{\small
Given information sources $S$ and $S'$ 
with possible worlds $\{D_1,\ldots,D_n\}$ and 
$\{D'_1,\ldots,D'_{n'}\}$ 
and tuple sets $T$ and $T'$ 

For every pair of possible-world relations $D_i \in S, D'_j \in S'$

\hs if $D_i$ and $D'_j$ are compatible then
let $Q_{ij} = D_i \cup D'_j$

end

The possible-worlds model of the result of integrating $S$ and $S'$ has the tuple set $T \cup T'$, 
and the set of possible-world
relations $Q_{ij}$ for every compatible pair $D_i$ and $D'_j$.
} 
\caption{Integration of uncertain data represented in the possible-worlds model}
\label{alg:integ-pw}
\end{algorithm}

\begin{example}
In Example~\ref{ex:scenario2}, the tuple sets for the two sources (Andy and Jane) are 
{\rm \{(Bob,CS100), (Bob,CS101), (Bob,CS102)\}} and 
{\rm \{(Bob,CS101), (Bob,CS102)\}}, respectively.
It is easy to verify that in this case the only compatible pair of possible-world relations are the second relation of Andy and the first relation of Jane
(See Figure~\ref{fig:andy-jane-1}). Hence, the integration result is {\rm \{(Bob,CS101)\}} as shown in Figure~\ref{fig:andy-jane-1}.

For case 2, the only difference is that the tuple set for Andy is {\rm \{(Bob,CS100), (Bob,CS101)\}}. Hence there are two pairs of compatible possible-world relations: In addition to second relation of Andy and first relation of Jane being compatible, we also have first relation of Andy compatible with second relation of Jane. This results in two possible-world relations in the integration: 
{\rm \{(Bob,CS101)\}} and {\rm \{(Bob,CS100), (Bob,CSC102)\}} as shown in Figure~\ref{fig:andy-jane-2}.
\examplebox
\end{example}

A logic-based approach to the representation and 
integration of uncertain data in the possible-world model 
was presented in \cite{Sadri:cikm12},
and shown to be equivalent to the superset-containment-based integration of \cite{Agrawal-DasSarma-Ullman-Widom:vldb10}.
It is easy to show the above algorithm is equivalent to the
logic-based and superset-containment-based integration.

\subsection{Compact Representation of Uncertain Data}
\label{sec:compact}

A number of models have been proposed for the representation of uncertain information
such as the ``maybe'' tuples model 
\cite{Codd:tods79,
Liu-Sunderraman:icde88,
LiuSun90,
LiuSun91},
set of alternatives or block-independent disjoint model (BID) 
\cite{BGP92,Benjelloun-etal:vldbj08,Dalvi-Suciu:pods07},
the probabilistic relation model 
\cite{Dalvi-Suciu:vldb04,Dalvi-Suciu:vldbj07},
and the U-relational database model 
\cite{Antova-Jansen-Koch-Olteanu:icde08}.
We have chosen the probabilistic relation model as a compact representation
of uncertain data for the integration of uncertain data \cite{Borhanian-Sadri:ideas13}.
Intuitively, 
this representation is based on the relational model
where each tuple $t$ is associated with a propositional logic formula $f(t)$
(called an {\em event} in \cite{Dalvi-Suciu:vldb04}.)
The Boolean variables in the formulas are called {\em event variables}. 
A probabilistic relation $r$ represents the set of 
possible-world relations corresponding to truth assignments 
to the set of event variables.
A truth assignment $\mu$ defines a possible-world relation
$r_{\mu} = \{t \mid t \in r {\mbox{ and }} f(t) = true {\mbox{ under $\mu$}}\}$.

\begin{example}
Probabilistic relations for the possible-worlds shown in 
Figure~\ref{fig:andy-jane-1} (Andy and Jane Case 1) are shown in 
Figure~\ref{fig:pr-for-andy-jane}.
\end{example}

\begin{figure}[h]
{\footnotesize
\begin{center}
\begin{tabular}{|c|c|c|} \hline
Bob     & CS100 & $x$\\ 
Bob     & CS101 & $\neg x$\\
Bob     & CS102 & {\em false}\\ \hline
\multicolumn{3}{c}{Andy}
\end{tabular}
\hs
\begin{tabular}{|c|c|c|} \hline
Bob     & CS101 & $y$\\ 
Bob     & CS102 & $\neg y$\\ \hline
\multicolumn{3}{c}{Jane}
\end{tabular}
\end{center}
\caption{Probabilistic relations of sources S1 (Andy)
and S2 (Jane) (Case 1)}
\label{fig:pr-for-andy-jane}
}
\end{figure}

\subsection{Integration of Uncertain Data Represented in the Probabilistic Relation Model}
\label{sec:integ-pr}

As mentioned earlier, for efficiency reasons 
a compact representation of uncertain data 
is utilized in practice. 
We will summarize an algorithm for the integration of uncertain
data represented in the probabilistic relation model
from \cite{Borhanian-Sadri:ideas13}.
First we need the following definition from 
\cite{Borhanian-Sadri:ideas13}.

\begin{definition}
\label{def-epr-relation}
An {\em extended probabilistic relation} is a probabilistic relation with a set of {\em event constraints}. Each event constraint is a propositional formula in event variables.
\end{definition} 

Semantics of an extended probabilistic relation is similar to that of probabilistic relation, with the exception that only truth assignments that satisfy event constraints are considered. More specifically, 
A truth assignment $\mu$ to event variables is {\em valid} if it satisfies all event constraints. A valid truth assignment $\mu$ defines a relation instance
$r_{\mu} = \{t \mid t \in r {\mbox{ and }} f(t) = true {\mbox{ under $\mu$}}\}$,
where $f(t)$ is the event formula associated with tuple $t$ in $r$.
The extended probabilistic relation $r$ represents the set of relations, called its
possible-world set, defined by the set of all valid truth assignments to the event variables.
We will use abbreviations {\em pr-relation} and 
{\em epr-relation} for probabilistic relation and extended probabilistic relation henceforth.

Given information sources $S$ and $S'$,
let $r$ and $r'$ be the pr-relations that represent the data in $S$ and $S'$, respectively. 
We represent a tuple in a pr-relation as $t@f$, 
where $t$ is the pure tuple, and $f$ is the 
propositional event formula associated with $t$.
Let $r = \{t_1@f_1,\ldots,t_n@f_n\}$, where $f_i$ is the event formula associated with the tuple $t_i$. 
Similarly, let $r' = \{u_1@g_1\ldots,u_m@g_m\}$. 
We assume the set of event variables of $r$ 
(\ie, event variables appearing in formulas $f_1,\ldots,f_n$) and those of $r'$ (\ie, event variables appearing 
in formulas $g_1,\ldots,g_m$) to be disjoint. 
If not, a simple renaming can be used to make the two sets disjoint.
$r$ and $r'$ can have zero or more common tuples. 
Assume, without loss of generality, 
that $r$ and $r'$ have $p$ tuples in common, 
$0 \le p \le min(n,m)$, $t_1 = u_1, \ldots, t_p = u_p$.
The integration algorithm is represented in
Algorithm~\ref{alg:integ-pr}.
In Algorithm~\ref{alg:integ-pr}, $f_i \equiv g_i$ is equivalent to the logical formula 
$(f_i \rightarrow g_i) \wedge (g_i \rightarrow f_i)$.
We will use the notation $q = r \uplus s$ to mean that $q$ is the epr-relation that is the result of integration of 
pr-relations $r$ and $s$.

\begin{algorithm}
{\small
Given information sources $S$ and $S'$,
let $r$ and $r'$ be the pr-relations that represent the data in $S$ and $S'$ as above. The result of integration of $S$ and $S'$ is represented by an epr-relation $q$ obtained as follows:
\begin{itemize}

\item
Copy to $q$ the tuples in $r$ that are not in common with $r'$, that is, $t_{p+1}@f_{p+1},\ldots,t_n@f_n$.

\item
Copy to $q$ the tuples in $r'$ that are not in common with $r$, that is, $u_{p+1}@g_{p+1},\ldots,u_m@g_m$.

\item
For each of the $p$ common tuples, 
copy to $q$ the tuple either from $r$ or from $r'$.

\item
For each of the $p$ common tuples, 
add a constraint $f_i \equiv g_i$, $i=1,\ldots,p$, to the set of event constraints of $q$.
\end{itemize}
} 
\caption{Integration of uncertain data represented by probabilistic relations}
\label{alg:integ-pr}
\end{algorithm}

It has been shown in
\cite{Borhanian-Sadri:ideas13}
that Algorithm~\ref{alg:integ-pr} is correct.
That is, when $q = r \uplus r'$ is obtained by this algorithm,
then the possible-worlds of $q$ coincide with the 
possible-worlds obtained by integrating possible-worlds of 
$r$ and $r'$ by Algorithm~\ref{alg:integ-pw}.

The complexity of Algorithm~\ref{alg:integ-pr} is 
$O(n \log n)$, where $n$ is
the size of input (pr-relations of the sources). 
While the complexity of the possible-worlds integration algorithm 
(Algorithm~\ref{alg:integ-pw})
is quadratic in the size of its input (possible-world relations of the sources) which itself can be exponential in the size 
of the input of Algorithm~\ref{alg:integ-pr}.

\begin{example}
\label{ex:andy-jane-pr-integration}
Consider the pr-relations for Andy and Jane shown in Figure~\ref{fig:pr-for-andy-jane}. We obtain the epr-relation of Figure~\ref{fig:epr-for-andy-jane} as the result of the integration.
There are two event constraints in this epr-relation, shown below the tuples.
The two sources have tuples {\rm (Bob,CS101)} and {\rm (Bob,CS102)} in common. The algorithm allows copying these tuples from either relation. In Figure~\ref{fig:epr-for-andy-jane} we have copied them from Jane's pr-relation. It is easy to verify that the only valid truth assignment to event variables for this epr-relation is $x =$ false, $y =$ true. 
The possible-world relation corresponding to this valid truth assignment contains one tuple, {\rm (Bob,CS101)} which is the same as the integration result shown in Figure~\ref{fig:andy-jane-1}.
\end{example}

\begin{figure}[h]
{\footnotesize
\begin{center}
\begin{tabular}{|c|c|c|} \hline
Bob     & CS100 & $x$\\ 
Bob     & CS101 & $y$\\
Bob     & CS102 & $\neg y$\\ \hline
\multicolumn{3}{|c|}{$\neg x \equiv y$} \\
\multicolumn{3}{|c|}{{\em false} $\equiv \neg y$} \\ \hline
\end{tabular}
\end{center}
\caption{Extended probabilistic relation of the integration 
of source S1 (Andy) and S2 (Jane)}
\label{fig:epr-for-andy-jane}
}
\end{figure}

\section{Integration of Probabilistic Uncertain Data}
\label{sec:prob-integ}

\subsection{Models of Probabilistic Uncertain Data}
\label{sec:models-of-prob-uncertain-data}

Both possible-worlds model and probabilistic relation model can be enhanced to represent probabilistic uncertain data:

\begin{definition}
\label{def:prob-uncertainDB}
A {\em probabilistic uncertain database} $U$ consists of 
a finite set of tuples $T(U)$ and a nonempty set of 
possible worlds $PW(U) = \{D_1, \ldots, D_n\}$, 
where each $D_i \subseteq T(U)$ is a certain database. 
Each possible world $D_i$ has a probability $0 < P(D_i) \le 1$ 
associated with it, such that
$\sum_{i=1}^n P(D_i) = 1$.
\end{definition}

A probabilistic relation can represent probabilistic uncertain database 
by associating probabilities with event variables.
Let $r = \{t1@f_1,\ldots,t_n@f_n\}$ be a pr-relation.
We can compute the probabilities associated 
with possible-world relations represented by $r$
as follows. 
Let $V = \{a_1, a_2, \ldots, a_k\}$ be the set of event variables of $r$.
Note that event variables are considered to be independent.
Let $\mu$ be a truth assignment 
to event variables. $\mu$ defines a relation instance
$r_{\mu} = \{t_i \mid t_i \in r {\mbox{ and }} f_i = true {\mbox{ under $\mu$}}\}$. The probability associated with $r_{\mu}$ is
\begin{equation}
\label{eq:pr-rel-probs}
\prod_{\mu(a_j) = true} P(a_j)
\prod_{\mu(a_j) = false} (1-P(a_j))
\end{equation}
A possible-world relation $r_i$ of $r$ can result from multiple truth assignments to event variables, in which case the probability of $r_i$, $P(r_i)$ is the sum of probabilities of $r_{\mu}$ for all truth assignments $\mu$ that generate $r_i$.

Our goal is to integrate information from sources 
containing probabilistic uncertain data, and
to compute the probability distribution of the 
possible-worlds of the result of the integration.
It has been shown that, in general, exact probabilities of the result of integration can not be obtained \cite{Sadri:cikm12}. Rather, only a {\em range} of probabilities can be computed
for each possible world in the integration. 
In this paper, we show that, 
under intuitive and reasonable assumptions, 
it is possible to obtain exact probabilities for the 
result of integration.

It is important to note that Equation~\ref{eq:pr-rel-probs}
is valid only when event variables are independent.
But we will see in the next section 
that this independence assumption no longer holds for
extended probabilistic relations. So, we are not
able to use Equation~\ref{eq:pr-rel-probs}
for epr-relations.

\subsection{Integration in the 
Probabilistic Possible-Worlds Framework}
\label{sec:relationship-to-pw}

A number of observations were made in \cite{Sadri:cikm12}
regarding integration of uncertain data represented in the
probabilistic possible-worlds model that are relevant to this work. We summarize these observations below.
 
Let $S$ and $S'$ be sources with possible worlds $\{D_1,\ldots,D_n\}$ and $\{D'_1,\ldots,D'_{n'}\}$, respectively. Consider the bi-partite graph $G$ defined by the relation 
$(D_i, D'_j)$: $D_i$ and $D'_j$ are compatible 
(See Definition~\ref{def:compatible} for compatible possible world relations). 
The graph $G$ is called the {\em compatibility graph} for sources $S$ and $S'$: There is an edge between $D_i$ and $D'_j$ if they are compatible. 
It has been shown that \cite{Sadri:cikm12}
\begin{itemize}

\item
Each connected component of $G$ is a complete bipartite graph.

\item
Let $H$ be a connected component of $G$. Then
\[
\sum_{D_i \in H} P(D_i) = \sum_{D'_j \in H} P(D'_j)
\]
These conditions have been called {\em probabilistic constraints} in \cite{Sadri:cikm12}. 

\end{itemize}

\begin{example}
\label{ex:pw-integ-1}
Consider the possible worlds of information sources $S$ and $S'$
shown in Figures~\ref{fig:probConstraints-S1}
and~\ref{fig:probConstraints-S2}.

\begin{figure}[h]
{\scriptsize
\begin{center}
\begin{tabular}{|c|c|} \hline
student & course \\ \hline
Bob     & CS100 \\ \hline
\multicolumn{2}{c}{D1}
\end{tabular}
\hs
\begin{tabular}{|c|c|} \hline
student & course \\ \hline
Bob     & CS100 \\
Bob     & CS101 \\ \hline
\multicolumn{2}{c}{D2}
\end{tabular}
\hs
\begin{tabular}{|c|c|} \hline
student & course \\ \hline
Bob     & CS101 \\ \hline
\multicolumn{2}{c}{D3}
\end{tabular}
\end{center}
\caption{Possible Worlds of source $S$}
\label{fig:probConstraints-S1}
} 
\end{figure}

\begin{figure}[h]
{\scriptsize
\begin{center}
\begin{tabular}{|c|c|} \hline
student & course \\ \hline
Bob     & CS100 \\ \hline
\multicolumn{2}{c}{D'1}
\end{tabular}
\hs
\begin{tabular}{|c|c|} \hline
student & course \\ \hline
Bob     & CS100 \\
Bob     & CS201 \\ \hline
\multicolumn{2}{c}{D'2}
\end{tabular}
\hs
\begin{tabular}{|c|c|} \hline
student & course \\ \hline
Bob     & CS201 \\ \hline
\multicolumn{2}{c}{D'3}
\end{tabular}
\hspace{10pt}
\begin{tabular}{|c|c|} \hline
student & course \\ \hline
Bob     & CS201 \\
Bob     & CS202 \\ \hline
\multicolumn{2}{c}{D'4}
\end{tabular}
\end{center}
\caption{Possible Worlds of source $S'$}
\label{fig:probConstraints-S2}
} 
\end{figure}

The compatibility bipartite graph $G$ for the possible-world relations of these sources is shown in 
Figure~\ref{fig:consistencyGraph}.
Note that we have 
$P(D_1) + P(D_2) = P(D'_1) + P(D'_2)$ and
$P(D_3) = P(D'_3) + P(D'_4)$
by the probabilistic constraints.
\examplebox

\begin{figure}[h]
\begin{center}
\includegraphics[width=12pc,height=8pc]{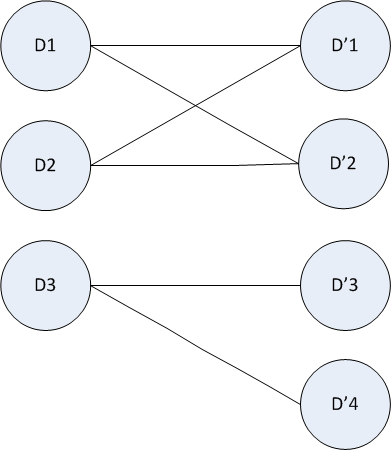}
\caption{Compatibility Graph for 
Example~\ref{ex:probConstraints}}
\label{fig:consistencyGraph}
\end{center}
\end{figure}
\end{example}

Let us concentrate on the top connected component portion of the compatibility bipartite graph $G$ shown in 
Figure~\ref{fig:consistencyGraph}. 
This connected component
gives rise to 4 possible worlds corresponding to
$D_1 \wedge D'_1$,
$D_1 \wedge D'_2$,
$D_2 \wedge D'_1$, and
$D_2 \wedge D'_2$.
We want to compute the probabilities of these possible-world relations, 
$P(D_1 \wedge D'_1)$,
$P(D_1 \wedge D'_2)$,
$P(D_2 \wedge D'_1)$, and
$P(D_2 \wedge D'_2)$,
given the probability distribution of the possible worlds of the sources, $P(D_1), P(D_2), P(D'_1), p(D'_2)$.

We have four unknowns. We can write the following four equations:

$P(D_1 \wedge D'_1) + P(D_1 \wedge D'_2) = P(D_1)$,

$P(D_2 \wedge D'_1) + P(D_2 \wedge D'_2) = P(D_2)$,

$P(D_1 \wedge D'_1) + P(D_2 \wedge D'_1) = P(D'_1)$,

$P(D_1 \wedge D'_2) + P(D_2 \wedge D'_2) = P(D'_2)$.

But, unfortunately, these equations are not independent.
Note that the probabilistic constraint requires that 
$P(D_1) + P(D_2) = P(D'_1) + P(D'_2)$. 
Hence, any one of the 4 equations can be obtained from the rest using the probabilistic constraint.
Hence we can only compute a probability range for each of these four possible-world relation.

So, how can we obtain exact probabilities for the possible-world relations of an integration? We make the following
{\em partial independence assumption}.

\subsection{Partial Independence Assumption}
\label{sec:partial-independence-pw}

{\em The only dependencies among the probabilities of 
possible-world relations are those induced by 
probabilistic constraints.}

Armed with this intuitive and reasonable assumption, we are able to compute exact probabilities for the result of an integration.

\begin{example}
\label{ex:pw-integ-2}
Consider again the top connected component in the 
compatibility graph of Example~\ref{ex:pw-integ-1}.
The structure of the graph tells us that if we have the evidence that the correct database of the first source $S$ is $D_1$, 
then we know the correct database of the second source $S'$ is
either $D'_1$ or $D'_2$. Similarly, 
if we have the evidence that the correct database of the first source $S$ is $D_2$,
then we know the correct database of the second source $S'$ is
either $D'_1$ or $D'_2$.
But, by the partial independence assumption, the knowledge of
$D_1$ or $D_2$ does not influence the probability of $D'_1$.
In other words, 
$P(D'_1 \mid D_1)$ is equal to 
$P(D'_1 \mid D_2)$.
Since $P(D'_1 \wedge D_1) = P(D'_1 \mid D_1) P(D_1)$ and
$P(D'_1 \wedge D_2) = P(D'_1 \mid D_2) P(D_2)$ we get

\[
\frac{P(D_1 \wedge D'_1)}{P(D_2 \wedge D'_1)} = 
\frac{P(D_1)}{P(D_2)}
\] 

\noindent
This serves as an additional equation that enables us to solve for the 4 unknowns. We get:

\vs
$P(D_1 \wedge D'_1) = P(D_1)P(D'_1)/(P(D_1)+P(D_2))$

$P(D_2 \wedge D'_1) = P(D_2)P(D'_1)/(P(D_1)+P(D_2))$

$P(D_1 \wedge D'_2) = P(D_1)P(D'_2)/(P(D_1)+P(D_2))$

$P(D_2 \wedge D'_2) = P(D_2)P(D'_2)/(P(D_1)+P(D_2))$
\end{example}

The observations of the above example can be generalized. Let $S_1$ and $S_2$ contain information in probabilistic possible-worlds model. Consider a connected component $G_1$ of the compatibility bipartite graph $G$ of $S_1$ and $S_2$. Let $D_1, \ldots, D_m$ and $D'_1,\ldots,D'_{m'}$ be the nodes of $G_1$ corresponding to possible worlds of $S_1$ and $S_2$, respectively. We can write the following $m + m'$ equations:

\[
\sum_{j=1}^{m'} P(D_i \wedge D'_j) = P(D_i), i = 1, \ldots, m
\]
and
\[
\sum_{i=1}^{m} P(D_i \wedge D'_j) = P(D'_j), j = 1, \ldots, m'
\]

But $m+m'-1$ of these equations are independent. Any one can be obtained from the rest using the probabilistic constraint
\[
\sum_{i=1}^{m} P(D_i) = \sum_{j=1}^{m'} P(D'_j)
\]
On the other hand, we have $m \times m'$ unknowns 
$P(D_i \wedge D'_j), i=1,\ldots,m, j=1,\ldots,m'$.
Additional equations are obtained from the independence assumption
\[
\frac{P(D_{1} \wedge D'_j)}{P(D_{i} \wedge D'_j)} =
\frac{P(D_{1})}{P(D_{i})}
\]
It can be shown that $(m-1) \times (m'-1)$ of these equations are independent. Together with the $m+m'-1$ equations of the first group we have the needed $m \times m'$ equations to solve for the unknowns. The solutions are,

\[
P(D_i \wedge D'_j) = 
\frac{P(D_i) P(D'_j)}{P}
\]
where $P$ is the probabilistic constraint constant
\[
P = \sum_{i=1}^{m} P(D_i) = \sum_{j=1}^{m'} P(D'_j)
\]

\section{Integration in the Probabilistic Relation Framework}
\label{sec:pr-framework}

In the previous section we presented an approach for 
the integration of porobabilistic uncertain data in 
the probabilistic possible-worlds framework. 
As mentioned earlier, the possible-worlds framework
is not suitable for practical applications. 
The size of the input, namely the possible-worlds relations,
can be exponential in the size of the equivalent representation
in the probabilistic relation framework. 
Further, we have a very efficient integration algorithm in the pr-relation framework. In this and next sections we concentrate
on the problem of determining the probability distribution for
the integration result in the pr-relation framework.

The integration algorithm in the pr-relation framework produces an extended pr-relation (Algorithm~\ref{alg:integ-pr}).
If the uncertain data is probabilistic, our task is to dtermine
the probability distribution for the result of the integration,
namely, an extended pr-relation. This problem was easy for
pure pr-relations: Event variables have probabilities associated with them, and probability distribution of the possible-worlds 
represneted by a pr-relation can be determined using the
independence assumption for event variables, as discussed in
Section~\ref{sec:models-of-prob-uncertain-data}.
But the independence assumption is no longer valid for 
extended pr-relaitons. 
Indeed, if we assume event variables are independent,
the sum of the probailities calculated for the possible-worlds 
of an epr-relation is not equal to 1. Intuitively, this is
due to the fact that only {\em valid} truth assignments,
those that satisfy the constraints, are taken into account.

The problem of determining the probabilities for epr-relations
in general remains open. But we will concentrate on the subclass
of interest, those epr-relations that can be obtained through
integration. In this section we present the subclass of
{\em integrated epr-relations} and present their properties.
Then in the next section we discuss how to determine the
probability distribution for this subclass.

This section contains discussions of theoretical nature, with
relatively long and complicated proofs of theorems. 
But we need this discussion to address 
{\em efficient} integration of 
probabilistic uncertain data in the pr-relation framework.
Proofs of the results in this section are presented 
in the Appendix.

\subsection{Integrated Extended Probabilistic Relations}
\label{sec:integrated-epr-relation}

While probability computation is straightforward for 
pr-relations, we do not have a general approach for 
probability computation for epr-relations. 
We will concentrate on a subclass of epr-relations: 
those that can be obtained as the result of integrating information sources. For data integration applications, 
this is the only class of epr-relations that are 
of interest to us. 

\begin{definition}
\label{integrated-epr-relation}
Given an extended probabilistic relation $q$,
we say $q$ is {\em integrated} if a pair of non-empty 
pr-relations $r$ and $s$ exists such that $q = r \uplus s$.
\end{definition}

First, we will present sufficient conditions for an epr-relation to be obtainable by integrating two information sources.

\begin{theorem}
\label{thm:integrated}
Let $q = \{t1@f_1,\ldots,t_n@f_n\}$ be an epr-relation,
with the set of event constraints $f_i \equiv g_i$, $i=1,\ldots,k$.
If a partition $(V, W)$ of event variables of $q$
exists such that
\begin{enumerate}

\item
For each tuple $t@f \in q$,
all event vaiables appearing in $f$ are in $V$ or all
are in $W$.

\item
For each event constraint $f \equiv g$ of $q$,
all event variables appearing in $f$ are in $V$ and
all event variables appearing in $g$ are in $W$, or vice versa.

\item
For each event constraint $f \equiv g$ of $q$,
there is a unique tuple $t$ such that $t@f \in q$
or $t@g \in q$.
\end{enumerate}
then $q$ is integrated.
\end{theorem}

\noindent
{\bf Proof.}
{\em Please see the Appendix}.

\subsection{Equivalence of pr-relation Pairs}
\label{sec:equivalence-of-pr-pairs}

For an integrated epr-relation $q$, there may exist multiple 
pr-relation pairs $(r_1, s_1)$, $(r_2,s_2), \ldots, (r_k, s_k)$
that can integrate to generate $q$. 
That is, $q = r_i \uplus s_i$, 
$i = 1, \ldots, k$. 
An example is presented in the Appendix
(Example~\ref{ex:multiple-pr-pairs}, 
see Figures \ref{fig:epr-for-multiple-pr-pairs},
\ref{fig:pr-pairs-1} and \ref{fig:pr-pairs-2}.)
We will show that all of these pairs are equivalent in probabilistic integration, in the sense that they generate exactly the same possible-worlds relations in the integration, with exactly the same probabilities. This result is quite important. It shows that the notion of integrated epr-relation is well-defined, in the sense that if an integrated epr-relation $q$ can be obtained by integrating alternative pr-relation pairs, all these integrations result in the same probabilistic uncertain database instance.
Our approach is as follows:

\begin{itemize}

\item
We associate a propositional formula with each possible world relation $r_i$ of a pr-relation or epr-relation $r$ that represents a probabilistic uncertain database. We call this formula the {\em event variable formula corresponding to the possible world relation $r_i$}.

\item
We give an alternative possible-worlds integration algorithm in terms of the event variable formulas associated with the possible-world relations of the two sources.

\item
We show that, for an epr-relation $q$ that satisfies the conditions of Theorem~\ref{thm:integrated} and all pr-relation pairs $(r,s)$ that generate $q$ by integration 
as obtained by Algorithm~\ref{alg:integrated} 
(presented in the Appendix), 
the event variable formulas of the possible-world relations 
of $q$ are equivalent. Hence, showing the equivalence of
possible-world relation set and their probability distribution. 
\end{itemize}  

\begin{definition}
\label{def:event-variable-formula}
Let $r = \{t_1@f_1, \ldots, t_n@f_n\}$ be a pr-relation, and let $T =\{t_1,\ldots,t_n\}$ be the tuple-set of $r$.
Consider a (regular) relation $r_i \subseteq T$.
The formula
\[\varphi_i = \bigwedge_{t_k \in r_i} f_k \bigwedge_{t_k \not \in r_i} \neg f_k\]
is called the {\em event-variable formula} of $r_i$.
\end{definition}

It is easy to verify the following observations:

\begin{itemize}
\item
Let $r$ be a pr-relation with tuple-set $T$, and
$r_i \subseteq T$ be a possible-world relation of $r$, $r_i \in PW(r)$.
Let $V$ be the set of event variables of $r$.
The event variable formula $\varphi_i$ is {\em true} for truth assignments to event variables in $V$ that yield
the possible world $r_i$ and {\em false} for all other  
truth assignments to event variables in $V$.

\item
Let $r$, $T$, and $V$ be as above. Consider a relation
$r_i \subseteq T$ that is not in the possible-world relations of $r$, $ r_i \not \in PW(r)$. Then the event variable formula $\varphi_i$ is a contradiction (that is, $\varphi_i$ is {\em false} for all truth assignments to event variables in $V$.)

\end{itemize}

We can extend the definition of event variable formulas 
for epr-relations, taking into account the event constraints.
The observations listed above hold for the following definition.

\begin{definition}
\label{def:epr-event-variable-formula}
Let $r = \{t_1@f_1, \ldots, t_n@f_n\}$ be an epr-relation. Let $T =\{t_1,\ldots,t_n\}$ be the tuple-set of $r$, and
$c_1, \ldots, c_m$ be the event constraints of $r$. Let
$C = c_1 \wedge \ldots \wedge c_m$.
Consider a (regular) relation $r_i \subseteq T$.
The formula
\[\varphi_i = C \wedge \bigwedge_{t_k \in r_i} f_k \bigwedge_{t_k \not \in r_i} \neg f_k \]
is called the {\em event-variable formula} of $r_i$.
\end{definition}

\subsubsection{Event-Variable Formulas for the Integration of pr-relations}
\label{sec:evf-integ}

Consider sources whose uncertain information is represented by pr-relations $r$ and $s$. 
We can show that event variable formulas of the possible-worlds relations of the integration of $r$ and $s$ can be obtained as the conjunction of the event variable formulas of the possible-worlds relations of $r$ and $s$. First we prove the following Lemma.

\begin{lemma}
\label{lem:formula-for-integration-part1}
Consider pr-relations
$r = \{t_1@f_1, \ldots, t_n@f_n\}$
and $s = \{u_1@g_1, \ldots, u_m@g_m\}$.
Let $V$ and $W$ be the set of event variable in $r$ and $s$, respectively. Without loss of generality, 
assume $V \cap W = \phi$. 
If not, a simple renaming can be used to make them disjoint.
Let $q = r \uplus s$ be the epr-relation obtained by 
the integration Algorithm~\ref{alg:integ-pr}. 
Let $r_i \in PW(r)$ and $s_j \in PW(s)$ with event-variable
formulas $\varphi_i$ and $\psi_j$, respectively.
Let $\xi = \varphi_i \wedge \psi_j$, and $\mu$ be a truth assignment
to variables in $V \cup W$. Then
if $r_i$ and $s_j$ are compatible, and if $\xi$ is {\em true}
under $\mu$,
then $\mu$ is a valid truth assignment for $q$. That is, 
all event constraints of $q$ are satisfied under $\mu$.
\end{lemma}

\negvs
\noindent
{\bf Proof.}
{\em Please see the Appendix}.

\begin{theorem}
\label{thm:formula-for-integration-part2}
Let $r$, $s$, $r_i$, $s_j$, $\varphi_i$, $\psi_j$, $\xi$, and $\mu$ be as defined 
in Lemma \ref{lem:formula-for-integration-part1}.
Then $\xi$ is the event-variable formula associated
with possible-world relation $q_{ij} = r_i \cup s_j$
of epr-relation $q = r \uplus s$.
\end{theorem}

\noindent
{\bf Proof.}
{\em Please see the Appendix}.

\subsubsection{Robustness Theorem}

Now we address the central problem of integrated epr-relations. Let $q$ be an epr-relation that satisfies the conditions of Theorem~\ref{thm:integrated}. Consider pr-relation pairs $(r,s)$ and $(r',s')$ obtained by Algorithms~\ref{alg:integrated} and~\ref{alg:partition} (presented in the Appendix)
such that $q = r \uplus s$ and $q = r' \uplus s'$. We will show that the event-variable formulas obtained for possible-world relations of $q$ through integration of $r$ and $s$ are equivalent to the event-variable formulas obtained through integration of $r'$ and $s'$.

\vs
First, we make a few observations about the partition algorithm, Algorithm~\ref{alg:partition}. 

\begin{itemize}

\item
The difference in alternative pr-relations $r$ and $r'$ (and $s$ and $s'$) obtained by Algorithm~\ref{alg:partition} come from the set of unlabeled nodes $X$. If $X$ is empty, then the algorithm generates a unique pr-relation pair. Otherwise, we are free to partition $X$ into $X_1$ and $X_2$ and add the corresponding event variables to $V_1$ and $W_1$.
As a result, we can obtain multiple pr-relation pairs.

\item
The edges in graph $H$ result from event constraints of $q$. 
If a node $A$ in $H$ does not have any incident edges, 
then none of the event variables represented by $A$ appear 
in an event constraint. 
Recall that the set $X$ consists exactly of these nodes
with no edges.

\item
Let $q = \{v_1@h_1,\ldots,v_n@h_n\}$, 
and assume event variables of some $h_i$ belong to a node $A$ in $X$. Note that all of the event variable in an $h_i$ should belong to the same node $A$ by Step 1 of the algorithm. Then the difference between $r$ and $r'$ (and $s$ and $s'$) correspond to such tuples $v_i@h_i$. That is, we may have 
$v_i@h_i \in r$, but $v_i@h_i \not \in r'$, while $v_i@h_i \not \in s$, but $v_i@h_i \in s'$. 
\end{itemize}

We will concentrate on the case where $r$ and $r'$ (and $s$ and $s'$) differ in a single tuple. We call this a {\em single-tuple} transformation. We show, for single-tuple transformation, the event-variable formulas generated for the possible-worlds relations of $q$ through integration of $r$ and $s$ are equivalent with the formulas generated through integration of $r'$ and $s'$. The general case, where $r$ and $r'$ (and, accordingly, $s$ and $s'$) differ in multiple tuples, can be obtained by multiple single-tuple transformations. The event-variable formulas for the possible-worlds relations of $q$ remain equivalent for each single-tuple transformation, and hence for the overall transformation.

\begin{theorem}
\label{thm:robustness}
Let $q$ be an epr-relation that satisfies the conditions of Theorem~\ref{thm:integrated}. Consider pr-relation pairs $(r,s)$ and $(r',s')$ obtained by Algorithms~\ref{alg:integrated} and~\ref{alg:partition} such that $q = r \uplus s$ and $q = r' \uplus s'$. Further, assume $r$ and $r'$ (accordingly, $s$ and $s'$) differ in a single tuple. Then the event-variable formulas obtained for possible-world relations of $q$ through integration of $r$ and $s$ are equivalent to the event-variable formulas obtained through integration of $r'$ and $s'$.
\end{theorem}

\noindent
{\bf Proof.}
{\em Please see the Appendix.}

\section{Integration in the Probabilistic Relation Framework 
-- Determining Probabilities}
\label{sec:prob-computation}

While probability computation is straightforward for 
pr-relations, we do not have a general approach for 
probability computation for epr-relations. 
The reason is that we can no longer assume event variables 
are independent. Event constraints impose certain 
dependencies among event variables.
In fact, it has been shown that
determining exact probabilities
of the result of integration is possible
only if we know the {\em correlation} between the sources. Otherwise, we can only obtain probability {\em ranges} \cite{Sadri:cikm12}.  
A similar observation has been noticed in the context of
probabilistic data exchange 
\cite{Fagin-Kimelfeld-Kolaitis:jacm11}.

We show that under an intuitive and reasonable assumption regarding the correlation of event variables of epr-relations 
we are able to compute the probabilities of the result of integration. 

\subsection{Partial Independence Assumption for
Extended Probabilistic Relations}
\label{sec:partial-independence-pr}

We make the following assumption:
{\em All event variables are independent except for the relationships induced by the event constraints.} 
In other words, the only correlations between event variables
are those resulting from event constraints. 

The following example demonstrates how this intuitive and reasonable assumption enables us to compute the probability distribution of an integration.

\begin{example}
\label{ex:probConstraints}
Consider the possible worlds of information sources $S$ and $S'$
from Example~\ref{ex:pw-integ-1},
shown in Figures~\ref{fig:probConstraints-S1}
and~\ref{fig:probConstraints-S2}.
Assume the probability distributions are
$P(D_1) = 0.3$,
$P(D_2) = 0.5$,
$P(D_3) = 0.2$,
$P(D'_1) = 0.35$,
$P(D'_2) = 0.45$,
$P(D'_3) = 0.05$, and
$P(D'_4) = 0.15$.


Algorithms for producing pr-relations for uncertain probabilistic databases have been presented in 
\cite{Borhanian-Sadri:ideas13,Dalvi-Suciu:vldbj07}.
We have used the algorithm of \cite{Borhanian-Sadri:ideas13}
to obtain the pr-relations $r_1$ and $r_2$ of 
Figure~\ref{fig:probRels-S1-S2-2} for the uncertain probabilistic database of 
Figures~\ref{fig:probConstraints-S1}
and~\ref{fig:probConstraints-S2}.
Probabilities of the event variables are also computed by the
algorithm and are:
$P(b_1) = 0.35$,
$P(b_2) = \frac{9}{13}$,
$P(b_3) = 0.25$,
$P(c_1) = 0.2$, and
$P(c_2) = 0.625$.

\begin{figure}[h]
{\footnotesize
\begin{center}
\begin{tabular}{|c|c|c|} \hline
student	& course	& $E$ \\ \hline
Bob		& CS100	& $\neg c_1$ \\
Bob		& CS101	& $c_1 \vee c_2$ \\ \hline
\multicolumn{3}{c}{pr-relation $r_1$}
\end{tabular}
\hs
\begin{tabular}{|c|c|c|} \hline
student	& course	& $E$ \\ \hline
Bob		& CS100	& $b_1 \vee b_2$ \\
Bob		& CS201	& $\neg b_1$ \\ 
Bob		& CS202	& $\neg b_1 \wedge \neg b_2 \wedge 
\neg b_3$ \\ \hline
\multicolumn{3}{c}{pr-relation $r_2$}
\end{tabular}
\end{center}
\caption{pr-relations for sources $S$ and $S'$}
\label{fig:probRels-S1-S2-2}
} 
\end{figure}

The result of integration is the epr-relation of 
Figure~\ref{fig:eprobRel-2}, obtained using 
Algorithm~\ref{alg:integ-pr}.
The possible-worlds relations of this epr-relation 
are shown in 
Figure~\ref{fig:probConstraints-S1-S2}.

\begin{figure}[h]
{\footnotesize
\begin{center}
\begin{tabular}{|c|c|c|} \hline
student	& course	& $E$ \\ \hline
Bob		& CS100	& $\neg c_1$ \\
Bob		& CS101	& $c_1 \vee c_2$ \\
Bob		& CS201	& $\neg b_1$ \\ 
Bob		& CS202	& $\neg b_1 \wedge \neg b_2 \wedge 
\neg b_3$ \\ \hline
\multicolumn{3}{|c|}{$\neg c_1 \equiv b_1 \vee b_2$} \\ \hline
\multicolumn{3}{c}{epr-relation $q = r_1 \uplus r_2$}
\end{tabular}
\end{center}
\caption{Extended Probabilistic relation for the integration of
sources $S$ and $S'$}
\label{fig:eprobRel-2}
} 
\end{figure}

\begin{figure}[h]
{\scriptsize
\begin{center}
\begin{tabular}{|c|c|} \hline
student & course \\ \hline
Bob     & CS100 \\ \hline
\multicolumn{2}{c}{(D1,D'1)}
\end{tabular}
\hs
\begin{tabular}{|c|c|} \hline
student & course \\ \hline
Bob     & CS100 \\
Bob     & CS201 \\ \hline
\multicolumn{2}{c}{(D1,D'2)}
\end{tabular}
\hs
\begin{tabular}{|c|c|} \hline
student & course \\ \hline
Bob     & CS100 \\
Bob     & CS101 \\ \hline
\multicolumn{2}{c}{(D2,D'1)}
\end{tabular}
\hspace{10pt}
\begin{tabular}{|c|c|} \hline
student & course \\ \hline
Bob     & CS100 \\
Bob     & CS101 \\
Bob     & CS201 \\ \hline
\multicolumn{2}{c}{(D2,D'2)}
\end{tabular}
\hs
\begin{tabular}{|c|c|} \hline
student & course \\ \hline
Bob     & CS101 \\
Bob     & CS201 \\ \hline
\multicolumn{2}{c}{(D3,D'3)}
\end{tabular}
\hs
\begin{tabular}{|c|c|} \hline
student & course \\ \hline
Bob     & CS101 \\
Bob     & CS201 \\
Bob     & CS202 \\ \hline
\multicolumn{2}{c}{(D3,D'4)}
\end{tabular}
\end{center}
\caption{Possible-world relations of the result of integration of sources $S$ and $S'$}
\label{fig:probConstraints-S1-S2}
} 
\end{figure}

How can we calculate the probability distribution of the result of integration (possible-world relations of 
Figure~\ref{fig:probConstraints-S1-S2})? 
The event-variable formulas for the 6 possible-world relations of the integration in this case are:

$\neg c_1 \wedge \neg c_2 \wedge b_1$

$\neg c_1 \wedge \neg c_2 \wedge \neg b_1 \wedge b_2$

$\neg c_1 \wedge c_2 \wedge b_1$

$\neg c_1 \wedge c_2 \wedge \neg b_1 \wedge b_2$

$c_1 \wedge \neg b_1 \wedge \neg b_2 \wedge b_3$

$c_1 \wedge \neg b_1 \wedge \neg b_2 \wedge b_3$

\noindent
By the partial independence assumption event variables are independent except for the relationships induced by the event constraints. The constraint $\neg c_1 \equiv b_1 \vee b_2$ induces a relationship between $c_1$ on one hand, and $b_1$ and $b_2$ on the other. The rest are still independent. So, for example, $c_1$ and $c_2$ are independent, and so are $c_2$ and $b_1$; etc... In particular, 
$b_1$ and $b_2$ are also independent. To compute the probability associated with an event-variable formula, we rewrite the formula so that it only contains mutually independent event variables. For example, $\neg c_1 \wedge \neg c_2 \wedge b_1$
is simplified to $\neg c_2 \wedge b_1$ using the equivalence $\neg c_1 \equiv b_1 \vee b_2$. Then we are able to compute the probabilities. 
In this example,
we obtain the following probabilities for the 
6 possible-world relations:
0.13125, 0.16875, 0.21875, 0.28125, 0.05, and 0.15.

Let us compare this approach with the integration in the 
probabilistic possible-worlds framework 
(Section~\ref{sec:partial-independence-pw}).
It is easy to verify that the probabilistic distribution
of the result of the integration computed by the formula
$P(D_i \wedge D'_j) = P(D_i)P(D_j)/P$ is exctly the same as the 
distribution obtained above. For example, the probability of
the possible world corresponding to $(D_1, D'_1)$ is
$0.3 \times 0.35 / (0.3 + 0.5) = 0.13125$.
\examplebox
\end{example}

\nop{
\begin{itemize}
\item
$(D_1,D'_1) = \{t_1\}$ corresponds to 1000, 1010, 1100, and 1110 (the vectors represent $(b_1 b_2 b_3 c_2)$.) Hence, the formula $b_1 \wedge \neg c_2$. The probability is 
$\frac{7}{20} \times (1-\frac{5}{8}) = \frac{21}{160} = 0.13125$

\item
$(D_1,D'_2) = \{t_1, t_3\}$ corresponds to 0100 and 0110. Hence, the formula $\neg b_1 \wedge b_2 \wedge \neg c_2$. The probability is 
$(1-\frac{7}{20}) \times \frac{9}{13} \times (1-\frac{5}{8}) = \frac{27}{160} = 0.16875$

\item
$(D_2,D'_1) = \{t_1, t_2\}$ corresponds to 1001, 1011, 1101, 1111. Hence, the formula $b_1 \wedge c_2$. The probability is 
$\frac{7}{20} \times \frac{5}{8} = \frac{35}{160} = 0.21875$

\item
$(D_2,D'_2) = \{t_1, t_2, t_3\}$ corresponds to 0101, 0111. Hence, the formula $\neg b_1 \wedge b_2 \wedge c_2$. The probability is 
$(1-\frac{7}{20}) \times \frac{9}{13} \times \frac{5}{8} = 
\frac{45}{160} = 0.28125$

\item
$(D_3,D'_3) = \{t_2, t_3\}$ corresponds to 0010, 0011. Hence, the formula $\neg b_1 \wedge \neg b_2 \wedge b_3$. The probability is 
$(1-\frac{7}{20}) \times (1-\frac{9}{13)} \times \frac{1}{4} = 
\frac{1}{20} = 0.05$

\item
$(D_3,D'_4) = \{t_2, t_3\}$ corresponds to 0000, 0001. Hence, the formula $\neg b_1 \wedge \neg b_2 \wedge \neg b_3$. The probability is 
$(1-\frac{7}{20}) \times (1-\frac{9}{13)} \times (1-\frac{1}{4}) = \frac{3}{20} = 0.15$
\end{itemize} 
} 

\subsection{Equivalence of Integration in the Two Frameworks}
\label{sec:equivalence}

We studied the problem of computing the probability distribution of the result of integration of probabilistic uncertain data using two main approaches: 
({\em i}) the probabilistic possible-worlds model approach and
({\em ii}) the probabilistic and extended probabilistic 
relation model approach.
The possible-worlds model is the accepted theoretical basis for uncertain data. But it is not practical for representation and integration due to exponential size. On the other hand, the probabilistic and extended probabilistic relation models are compact and highly efficient approaches to probabilistic uncertain information representation and integration.

\begin{itemize}

\item
In the first approach, we made the partial-independence assumption that the only dependencies among the possible-worlds of the sources are those induced by probabilistic constraints. Using this assumption, we could obtain a relatively simple formula for the computation of probabilities for the result of integration.

\item
In the second approach, we made the partial-independence assumption that the only dependencies among event variables of the pr-relations are those induced by event constraints. Using this assumption, we could obtain probabilities for the result of integration. Our results regarding different (but equivalent) pr-relation pairs for the sources play a key role in making the probability computation possible.

\item
The two approaches are closely related. In fact, event constraints of the epr-relation that represents the result of integration enforce the probabilistic constraints on the possible-worlds of the sources. The independence assumption regarding possible-worlds of the two sources, except only when induced by probabilistic constraints, is closely related to the independence assumption regarding the event variables of the two pr-relations, except only when induced by the event constraints. We can consider the two approaches equivalent, except they operate in different frameworks, one in the possible-worlds framework, the other in the pr-relations framework.

\item
The important difference in the two approaches is the efficiency: While the possible-worlds framework is not practical for integration due to exponential size, the pr- and epr-relation framework is a compact and highly efficient approach to probabilistic uncertain information representation and integration.

\end{itemize}

\section{Conclusion}
\label{sec:conclusion}

We focused on data integration from sources containing probabilistic uncertain information, in particular, on computing the probability distribution of the result of integration. We presented integration algorithms for data represented in two frameworks: The probabilistic possible-worlds model and the probabilistic relation model. 
In the latter case the result of integration is represented 
by an extended probabilistic relation. 
We introduced an important subclass of this extended model, namely, those epr-relations that result from integration 
of uncertain information. 
Alternative approaches to the computation of the probability 
distribution were presented in the two frameworks, 
and shown to be equivalent.


\newcommand{\sigmod}{Proceedings of ACM SIGMOD International
Conference on Management of Data}

\newcommand{\pvldb}{Proceedings of the VLDB Endowment}

\newcommand{\icde}{Proceedings of IEEE International Conference on Data
Engineering}

\newcommand{\vldbj}{The VLDB Journal}

\newcommand{\ideas}{Proceedings of International Database
Engineering and Applications, IDEAS}

\newcommand{\cacm}{Communications of the ACM}

\newcommand{\vldb}{Proceedings of International Conference on 
Very Large Databases}

\newcommand{\icdt}{Proceedings of International Conference on Database Theory}

\newcommand{\cikm}{Proceedings of International Conference on Information
and Knowledge Management}

\newcommand{\tkde}{IEEE Transactions on Knowledge and Data Engineering}

\newcommand{\tods}{ACM Transactions on Database Systems}

\newcommand{\pods}{Proceedings of ACM Symposium on Principles of 
Database Systems}

\newcommand{\jacm}{Journal of the ACM}

\newcommand{\mud}{Proceedings of VLDB Workshop on
Management of Uncertain Data}

\newcommand{\jdiq}{ACM Journal of Data and Information Quality}

\newpage
\section*{Appendix}
\label{sec:appendix}

{\bf Proof of Theorem \ref{thm:integrated}.}

We will show that
if conditions of Theorem~\ref{thm:integrated} hold,
Algorithm~\ref{alg:integrated} can be used to produce pr-relations $r$ and $s$
such that $q \equiv r \uplus s$.
Step 1 of the algorithm partitions the tuples of $q$ onto
pr-relations $r$ and $s$.
By condition 1 of Theorem~\ref{thm:integrated},
this partition is well-defined.
Step 2 of the algorithm adds more tuples to $r$ and/or $s$
to complete the construction.

\begin{algorithm}
{\small
Let $E(f)$ represent the set of
event variables of a formula $f$.
\begin{enumerate}

\item
Partition tuples of $q$ as follows.
Let $r = \{t@f \in q \mid E(f) \subseteq V\}$
and $s = \{t@f \in q \mid E(f) \subseteq W\}$.

\item
For each constraint $f \equiv g$ of $q$,
if $t@f \in r$ (or $t@f \in s$), then add $t@g$ to $s$ 
(or to $r$), or
if $t@g \in r$ (or $t@g \in s$), then add $t@f$ to $s$ 
(or to $r$).
\end{enumerate}
\caption{Obtaining pr-relations $r$ and $s$ 
such that $q = r \uplus s$}
\label{alg:integrated}
} 

\end{algorithm}

Next, we should show that given epr-relation $q$,
pr-relations $r$ and $s$ produced by Algorithm~\ref{alg:integrated}
satisfy $r \uplus s \equiv q$. 
Assume $r \uplus s = q'$. 
We will first verify that $q'$ has the same set of event constraints as $q$.
For each constraint $f \equiv g$ in $q$, by Conditions 2 and 3 of the theorem,
there is a unique tuple $t@f$ or $t@g$ in $q$.
Hence, by step 1 of the construction algorithm, 
$t@f$ or $t@g$ is in $r$ or $s$.
Without loss of generality, assume $t@f \in r$. 
Step 2 of the construction algorithm
adds $t@g$ to $s$. Then the integration algorithm 
(Algorithm~\ref{alg:integ-pr})
generates $f \equiv g$ for $q' = r \uplus s$.

Finally, we should show that $q'$ has the same (or equivalent) set of tuples as $q$.
By Algorithm~\ref{alg:integrated},
for all $t@f \in q$, either $t@f \in r$ or $t@f \in s$.
Then, by the integration algorithm, either $t@f \in q'$ or $t@g \in q'$ for some $g$ that is equivalent to $f$,
$g \equiv f$. It follows that set of tuples of $q'$ and $q$ are equivalent.
Example~\ref{ex:multiple-pr-pairs} given further below demonstrates Algorithm~\ref{alg:integrated}.

Given an epr-relation $q$ how can we determine whether it satisfies the conditions of 
Theorem~\ref{thm:integrated}?  Condition 3 of the theorem can be checked easily. Algorithm~\ref{alg:partition}, presented below, can be used to determine if an epr-relation $q$ satisfies conditions 1 and 2, and also produce the partitions $(V, W)$ of event variables of $q$ according to
Theorem~\ref{thm:integrated}.

\begin{algorithm}

Given extended probabilistic relation $q$ 
with constraints $f_i \equiv g_i$, $i=1,\ldots,p$, and
tuples $\{v_1@h_1, \ldots, v_l@h_l\}$,
let $E(q)$ be the set of event variables appearing in $q$. 

\BlankLine

{\bf Initialization}:
For each event variable $e \in E(q)$, 
construct a (singleton) event-variable set containing $e$.

\vs
{\bf Step 1}

\ForEach{$h_i$, $i=1,\ldots,l$ and each pair $(e,e')$ of event variables appearing in $h_i$}{
  Let $A$ and $A'$ be the event variable sets containing $e$ and $e'$
  (that is, $e \in A$ and $e' \in A'$)\;
  Replace $A$ and $A'$ with $A \cup A'$.\;
  }

Do the same for each $f_i$ and for each $g_i$, $i=1,\ldots,p$.

Let $A_1, A_2, \ldots$ be the sets of event variables obtained.

\vs
{\bf Step 2}

Construct a graph $H$ as follows:
Nodes of $H$ correspond to $A_1, A_2, \ldots$.
There is an edge between $A_i$ and $A_j$ if there are
event variables $e \in A_i$ and $e' \in A_j$ and
$q$ has a constraint $f \equiv g$ with $e \in E(f)$
and $e' \in E(g)$, or vice-versa. 

\Repeat{all nodes in connected components are labeled}{
Start with a (randomly chosen) node $A$ in a (randomly chosen) connected component of $H$ and label it $V$

\ForEach{Node $A$}{
  if $A$ is labeled $V$ then label all nodes connected to 
$A$ by $W$;

  if $A$ is labeled $W$ then label all nodes connected to 
$A$ by $V$
}
}

If a node is labeled both $V$ and $W$ then return failure: $q$ does not satisfy the conditions of Theorem~\ref{thm:integrated}. Else

Let $V_1$ be the set of event variables of all nodes 
labeled $V$, $W_1$ be the set of event variables of all nodes 
labeled $W$, and $X$ be the set of unlabeled nodes (if any).
Partition $X$ into $X_1$ and $X_2$ randomly. 
Let $Y_1$ be the set of event variables in 
$X_1$ nodes, and $Y_2$ be the set of event variables 
in $X_2$ nodes. 

Let $V = V_1 \cup Y_1$ and $W = W_1 \cup Y_2$.
\caption{Partition}
\label{alg:partition}
\end{algorithm}

Algorithm~\ref{alg:partition} works in two steps:
In the first step, sets of event variables that should appear together (in $V$ or in $W$) are identified. At the end of this step, each set $A_1, A_2, \ldots$ contains a set of event variables that must appear together.

In the second step, it is determined whether it is possible to combine the event-variable sets of step 1
into the partitions $V$ and $W$ that satisfy condition 2 of Theorem~\ref{thm:integrated}. We construct a graph $H$ where each node represents a set of event variables $A_i$ from Step 1. If a node $A_i$ is connected to $A_j$ in $H$, then the event variables of $A_i$ and those of $A_j$ must belong to different partitions. The algorithm labels nodes in the connected components of $H$. If the node $A_i$ is connected to $A_j$, they are labeled by different partitions. The labelling fails if a node must be labeled both $V$ and $W$. Otherwise it succeeds. 
At the end of labeling, nodes that do not have an incident edge remain unlabeled. Event variables represented by these nodes are free to be included in $V$ or in $W$. As a result, multiple $(V,W)$ pairs are possible as shown in Example~\ref{ex:multiple-pr-pairs}.

The complexities of Algorithms~\ref{alg:integrated} 
and~\ref{alg:partition} are linear as each tuple of the input epr-relation is examined once.

\begin{example}
\label{ex:multiple-pr-pairs}
Consider the epr-relation of 
Figure~\ref{fig:epr-for-multiple-pr-pairs}.
Step 1 of Algorithm~\ref{alg:partition} produces event variable sets $\{a\}$, $\{b\}$, and $\{c,d\}$. The graph $H$ of step 2 has only one edge between (nodes representing)
$\{a\}$ and $\{c,d\}$. Step 2 labels $\{a\}$ with $V$ and 
$\{c,d\}$ with $W$, while $\{b\}$ remains unlabeled. Hence, there are two ways to obtain the partition for 
Theorem~\ref{thm:integrated}: by combining $\{b\}$ with $V$; or by combining $\{b\}$ with $W$. We obtain the two pairs 
$V_1 = \{a,b\}$, $W_1 = \{c,d\}$; and $V_2 = \{a\}$, 
$W_2 = \{b,c,d\}$. The resulting pr-relation pairs whose integration generates the epr-relation of  
Figure~\ref{fig:epr-for-multiple-pr-pairs}
are shown in
Figures~\ref{fig:pr-pairs-1} and~\ref{fig:pr-pairs-2}.
\end{example}

\begin{figure}[h]
{\footnotesize
\begin{center}
\begin{tabular}{|c|c|} \hline
Tuple     & $E$ \\ \hline
$t_1$     & $a$ \\
$t_2$     & $b$ \\
$t_3$     & $\neg c \vee d$ \\ \hline
\multicolumn{2}{|c|}{$a \equiv c$} \\ \hline
\end{tabular}
\end{center}
\caption{epr-relation for 
Example~\ref{ex:multiple-pr-pairs}}
\label{fig:epr-for-multiple-pr-pairs}
} 
\end{figure}

\begin{figure}[h]
{\footnotesize
\begin{center}
\begin{tabular}{|c|c|} \hline
Tuple     & $E$ \\ \hline
$t_1$     & $a$ \\
$t_2$     & $b$ \\ \hline
\end{tabular}
\hs
\begin{tabular}{|c|c|} \hline
Tuple     & $E$ \\ \hline
$t_1$     & $c$ \\
$t_3$     & $\neg c \vee d$ \\ \hline
\end{tabular}
\end{center}
\negvs
\negvs
\caption{pr-relation pair for 
Example~\ref{ex:multiple-pr-pairs}}
\label{fig:pr-pairs-1}
} 
\end{figure}

\begin{figure}[h]
{\footnotesize
\begin{center}
\begin{tabular}{|c|c|} \hline
Tuple     & $E$ \\ \hline
$t_1$     & $a$ \\ \hline
\end{tabular}
\hs
\begin{tabular}{|c|c|} \hline
Tuple     & $E$ \\ \hline
$t_1$     & $c$ \\
$t_2$     & $b$ \\
$t_3$     & $\neg c \vee d$ \\ \hline
\end{tabular}
\end{center}
\caption{Alternative pr-relation pair for 
Example~\ref{ex:multiple-pr-pairs}}
\label{fig:pr-pairs-2}
} 
\end{figure}

\noindent
{\bf Proof of Lemma \ref{lem:formula-for-integration-part1}.}

Assume, without loss of generality, that $r$ and $s$ have $p$ common regular tuples
$t_k = u_k$, $k=1,\ldots,p$. 
Then $q = r \uplus s$ has $p$ event constraints
$f_k \equiv g_k$, $k=1,\ldots,p$.
Since $r_i$ and $s_j$ are compatible, then there is no tuple
$t_k \in T(r) \cap T(s)$ such that $t_k \in r_i$ and $t_k \not \in s_j$ or vice versa.
Further, since $\xi$ is {\em true}, then $\varphi_i$ and $\psi_j$ are {\em true}.
It follows that, for all $t_k \in T(r) \cap T(s)$,
either 
(1) $t_k \in r_i$ and $t_k \in s_j$ and hence
both $f_k$ and $g_k$ are {\em true} under 
truth assignment $\mu$, or 
(2) $t_k \not \in r_i$ and $t_k \not \in s_j$ and hence
$f_k$ and $g_k$ are both {\em false} under 
truth assignment $\mu$.
Hence, all event constraints $f_k \equiv g_k$, $i=1,\ldots,p$, 
are satisfied under $\mu$.

\vsp
\noindent
{\bf Proof of Theorem \ref{thm:formula-for-integration-part2}.}

Consider the truth assignment $\mu$ to event variables 
$V \cup W$.
By Lemma \ref{lem:formula-for-integration-part1}, if $\xi$
is {\em true} under $\mu$, then $\mu$ is legal.
Further, if $\xi$ is {\em true} under $\mu$, then so are $\varphi_i$ and $\psi_j$.
Hence, $f_k$ is {\em true} for all tuples $t_k \in r_i$, and it is {\em false}
for all tuples $t_k \in (T(r) - r_i)$.
Similarly, $g_k$ is {\em true} for all tuples $u_k \in s_j$, and it is {\em false}
for all tuples $u_k \in (T(s) - s_j)$.
Consider a tuple $v@h \in q = r \uplus s$.
(Note that $v@h$ is either $t_k@f_k$ or $u_k@g_k$ by the integration Algorithm~\ref{alg:integ-pr}.)
It is easy to see that $h$ is {\em true} under $\mu$ if and only if $v \in r_i \cup s_j$.
It follows that $\xi$ is {\em true} for all valid truth assignments that
yield the possible world $q_{ij} = r_i \cup s_j$, and, hence, $\xi$ is (equivalent to) the event-variable formula for
$q_{ij}$.

\vsp
\noindent
{\bf Proof of Theorem \ref{thm:robustness}.}

Let 
$r = \{t_1@f_1, \ldots, t_n@f_n\}$ and 
$s = \{u_q@g_1, \ldots, u_m@g_m\}$. Let $r_i \in PW(r)$ be compatible with $s_j \in PW(s)$. So, $q$ will have a possible-world relation $q_{ij} = r_i \cup s_j$. Let event-variable formulas for $r_i$ and $s_j$ be $\varphi_i$ and $\psi_j$, respectively. Hence, the event-variable for $q_{ij}$ is $\psi = \varphi_i \wedge \psi_j$ as shown in Section~\ref{sec:evf-integ}.

Now consider the alternative pr-relation pair $r'$ and $s'$ that differ from $r$ and $s$ in a single tuple, say $v@h$. That is, $r$ contains $v@h$ but $r'$ does not, while $s$ does not contain $v@h$ but $s'$ does. Let's consider how $q_{ij}$ is obtained in the integration of $r'$ and $s'$. We should have $r'_i \in PW(r')$ and $s'_j \in PW (s')$ that are compatible, and $q_{ij} = r'_i \cup s'_j$. We distinguish two cases, $v \in q_{ij}$ and $v \not \in q_{ij}$.

\vs
\noindent
{\bf Case 1:} $v \in q_{ij}$. In this case $v \in r_i$, $v \not \in s_j$ while $v \not \in r'_i$, $v \in s'_j$. The difference between event-variable formulas $\varphi_i$ (for $r_i$) and $\varphi'_i$ (for $r'i$) is only in the conjunct $h$: $\varphi_i$ has the conjunct, but $\varphi'_i$ does not. In other words,
$\varphi_i = \varphi'_i \wedge h$. Similarly, we have
$\psi'_j = \psi_j \wedge h$, for $s'_j$ and $s_j$. It follows that the event-variable formulas for $q_{ij}$ obtained by integrating $r$ and $s$, namely,
$\varphi_i \wedge \psi_j$ is equivalent to the formula obtained by integrating $r'$ and $s'$, namely,
$\varphi'_i \wedge \psi'_j$. 

\vs
\noindent
{\bf Case 2:} $v \not \in q_{ij}$. In this case $v$ is not in any of $r_i$, $s_j$, $r'_i$, nor $s'_j$.
The difference between event-variable formulas $\varphi_i$ (for $r_i$) and $\varphi'_i$ (for $r'i$) is only in the conjunct $\neg h$: $\varphi_i$ has the conjunct, but $\varphi'_i$ does not. This is due to the fact that $v$ is in the tuple-set of $q$, but it is not in $r_i$. So, $\varphi_i$ contains the conjunct $\neg h$. On the other hand, $v$ is not in the tuple-set of $r'_i$. So, $\varphi'_i$ does not contain the conjunct.
Similarly, $\psi'_j$ contains the conjunct $\neg h$, while $\psi_j$ does not. Again, it follows that the event-variable formulas for $q_{ij}$ obtained by integrating $r$ and $s$, namely,
$\varphi_i \wedge \psi_j$ is equivalent to the formula obtained by integrating $r'$ and $s'$, namely,
$\varphi'_i \wedge \psi'_j$. 


\begin{thebibliography}{10}

{\small
\bibitem{Abiteboul-Kanellakis-Grahne:sigmod87}
Serge Abiteboul, Paris~C. Kanellakis, and G{\"o}sta Grahne.
\newblock On the representation and querying of sets of possible worlds.
\newblock In {\em \sigmod}, pages 34--48, 1987.

\bibitem{Agrawal-DasSarma-Ullman-Widom:vldb10}
Parag Agrawal, Anish~Das Sarma, Jeffrey~D. Ullman, and Jennifer Widom.
\newblock Foundations of uncertain-data integration.
\newblock {\em \pvldb}, 3(1):1080--1090, 2010.

\bibitem{Antova-Jansen-Koch-Olteanu:icde08}
Lyublena Antova, Thomas Jansen, Christoph Koch, and Dan Olteanu.
\newblock Fast and simple relational processing of uncertain data.
\newblock In {\em \icde}, pages 983--992, 2008.

\bibitem{Antova-Koch-Olteanu:icde07}
Lyublena Antova, Christoph Koch, and Dan Olteanu.
\newblock 10$^{\mbox{10$^{\mbox{6}}$}}$ worlds and beyond: Efficient
  representation and processing of incomplete information.
\newblock In {\em \icde}, pages 606--615, 2007.

\bibitem{BGP92}
Daniel Barbar{\'a}, Hector Garcia-Molina, and Daryl Porter.
\newblock The management of probabilistic data.
\newblock {\em \tkde}, 4(5):487--502, October 1992.

\bibitem{Benjelloun-etal:vldbj08}
Omar Benjelloun, Anish~Das Sarma, Alon~Y. Halevy, Martin Theobald, and Jennifer
  Widom.
\newblock Databases with uncertainty and lineage.
\newblock {\em \vldbj}, 17(2):243--264, 2008.

\bibitem{Borhanian-Sadri:ideas13}
Amir~Dayyan Borhanian and Fereidoon Sadri.
\newblock A compact representation for efficient uncertain-information
  integration.
\newblock In {\em \ideas}, pages 122--131, 2013.

\bibitem{Chen-Chirkova-Sadri-Salo:actainformatica13}
Dongfeng Chen, Rada Chirkova, Fereidoon Sadri, and Tiia~J. Salo.
\newblock Query optimization in information integration.
\newblock {\em Acta Informatica}, 50(4):257--287, 2013.

\bibitem{Codd:tods79}
E.~F. Codd.
\newblock Extending the database relational model to capture more meaning.
\newblock {\em \tods}, 4(4):397--434, December 1979.

\bibitem{Dalvi-Re-Suciu:cacm09}
Nilesh~N. Dalvi, Christopher R{\'e}, and Dan Suciu.
\newblock Probabilistic databases: diamonds in the dirt.
\newblock {\em \cacm}, 52(7):86--94, 2009.

\bibitem{Dalvi-Suciu:vldb04}
Nilesh~N. Dalvi and Dan Suciu.
\newblock Efficient query evaluation on probabilistic databases.
\newblock In {\em \vldb}, pages 864--875, 2004.

\bibitem{Dalvi-Suciu:vldbj07}
Nilesh~N. Dalvi and Dan Suciu.
\newblock Efficient query evaluation on probabilistic databases.
\newblock {\em \vldbj}, 16(4):523--544, 2007.

\bibitem{Dalvi-Suciu:pods07}
Nilesh~N. Dalvi and Dan Suciu.
\newblock Management of probabilistic data: foundations and challenges.
\newblock In {\em \pods}, pages 1--12, 2007.

\bibitem{Dong-Halevy-Yu:vldb07}
Xin~Luna Dong, Alon Halevy, and Cong Yu.
\newblock Data integration with uncertainty.
\newblock In {\em \vldb}, pages 687--698, 2007.

\bibitem{Dong-Halevy-Yu:vldbj09}
Xin~Luna Dong, Alon~Y. Halevy, and Cong Yu.
\newblock Data integration with uncertainty.
\newblock {\em \vldbj}, 18(2):469--500, 2009.

\bibitem{Eshmawi-Sadri:ideas09}
Ala~A. Eshmawi and Fereidoon Sadri.
\newblock Information integration with uncertainty.
\newblock In {\em \ideas}, pages 284--291, 2009.

\bibitem{Fagin-Kimelfeld-Kolaitis:jacm11}
Ronald Fagin, Benny Kimelfeld, and Phokion~G. Kolaitis.
\newblock Probabilistic data exchange.
\newblock {\em \jacm}, 58(4):15, 2011.

\bibitem{Haas:icdt07}
Laura~M. Haas.
\newblock Beauty and the beast: The theory and practice of information
  integration.
\newblock In {\em \icdt}, pages 28--43, 2007.

\bibitem{Halevy-etal:sigmod05}
Alon~Y. Halevy, Naveen Ashish, Dina Bitton, Michael~J. Carey, Denise Draper,
  Jeff Pollock, Arnon Rosenthal, and Vishal Sikka.
\newblock Enterprise information integration: successes, challenges and
  controversies.
\newblock In {\em \sigmod}, pages 778--787, 2005.

\bibitem{Halevy-Rajaraman-Ordille:vldb06}
Alon~Y. Halevy, Anand Rajaraman, and Joann~J. Ordille.
\newblock Data integration: The teenage years.
\newblock In {\em \vldb}, pages 9--16, 2006.

\bibitem{Liu-Sunderraman:icde88}
K.~C. Liu and R.~Sunderraman.
\newblock On representing indefinite and maybe information in relational
  databases.
\newblock In {\em \icde}, pages 250--257, 1988.

\bibitem{LiuSun90}
K.~C. Liu and R.~Sunderraman.
\newblock Indefinite and maybe information in relational databases.
\newblock {\em \tods}, 15(1):1--39, March 1990.

\bibitem{LiuSun91}
K.~C. Liu and R.~Sunderraman.
\newblock A generalized relational model for indefinite and maybe information.
\newblock {\em \tkde}, 3(1):65--77, March 1991.

\bibitem{Magnani-Montesi:mud07}
Matteo Magnani and Danilo Montesi.
\newblock Uncertainty in data integration: current approaches and open
  problems.
\newblock In {\em \mud}, pages 18--32, 2007.

\bibitem{Magnani-Montesi:jdiq10}
Matteo Magnani and Danilo Montesi.
\newblock A survey on uncertainty management in data integration.
\newblock {\em \jdiq}, 2(1), 2010.

\bibitem{Olteanu-Huang-Koch:icde09}
Dan Olteanu, Jiewen Huang, and Christoph Koch.
\newblock {SPROUT}: Lazy vs. eager query plans for tuple-independent
  probabilistic databases.
\newblock In {\em \icde}, pages 640--651, 2009.

\bibitem{Re-Dalvi-Suciu:icde07}
Christopher Re, Nilesh~N. Dalvi, and Dan Suciu.
\newblock Efficient top-k query evaluation on probabilistic data.
\newblock In {\em \icde}, pages 886--895, 2007.

\bibitem{Sadri:cikm12}
Fereidoon Sadri.
\newblock On the foundations of probabilistic information integration.
\newblock In {\em \cikm}, pages 882--891, 2012.

\bibitem{Sarma-et-al:vldbj09}
Anish~Das Sarma, Omar Benjelloun, Alon~Y. Halevy, Shubha~U. Nabar, and Jennifer
  Widom.
\newblock Representing uncertain data: models, properties, and algorithms.
\newblock {\em \vldbj}, 18(5):989--1019, 2009.

\bibitem{Sarma-Benjelloun-Halevy-Widom:icde06}
Anish~Das Sarma, Omar Benjelloun, Alon~Y. Halevy, and Jennifer Widom.
\newblock Working models for uncertain data.
\newblock In {\em \icde}, page~7, 2006.

\bibitem{Sen-Deshpande:icde07}
Prithviraj Sen and Amol Deshpande.
\newblock Representing and querying correlated tuples in probabilistic
  databases.
\newblock In {\em \icde}, pages 596--605, 2007.
}

\end{thebibliography}
\end{document}